# Graphene-Graphite *Quilts* for Thermal Management of High-Power GaN Transistors

Zhong Yan, Guanxiong Liu, Javed M. Khan[x] and Alexander A. Balandin*

Nano-Device Laboratory, Department of Electrical Engineering and Materials Science and Engineering Program, University of California, Riverside, California 92521 USA

[x] Current Affiliation: Intel Corporation, Hillsborough, Oregon, USA

[*] Corresponding author: balandin@ee.ucr.edu

**Abstract**

Self-heating is a severe problem for high-power GaN electronic and optoelectronic devices. Various thermal management solutions, e.g. flip-chip bonding or composite substrates have been attempted. However, temperature rise still limits applications of the nitride-based technology. Here we demonstrate that thermal management of GaN transistors can be substantially improved via introduction of the alternative heat-escaping channels implemented with few-layer graphene – an excellent heat conductor. We have transferred few-layer graphene to AlGaN/GaN heterostructure field-effect transistors on SiC substrates to form the "graphene-graphite quilts" – lateral heat spreaders, which remove heat from the channel regions. Using the micro-Raman spectroscopy for *in-situ* monitoring we have shown that temperature can be lowered by as much as ~ 20$^{o}$C in such devices operating at ~13-W/mm power density. The simulations suggest that the efficiency of the "graphene quilts" can be made even higher in GaN devices on thermally resistive sapphire substrates and in the designs with the closely located heat sinks. Our results open a novel application niche for few-layer graphene in high-power electronics.

Self-heating is a severe problem for the high-power GaN electronic, optoelectronic and photonic devices [1-3]. Temperature rise and non-uniform distribution of dissipated power in GaN transistors result in degradation of the drain current, gain and output power, as well as an increase in the gate-leakage current and poor reliability [1]. Various thermal management solutions, e.g. flip-chip (FC) bonding [4] or diamond composite substrates [5] have been attempted. However, the hot spots, which appear owing to the high power densities and relatively high thermal resistance of the substrates [6-7], still limit applications of the nitride-based technology [1-3]. For example, AlGaN/GaN HFETs are attractive devices for high-frequency high-power communications and radar applications [1-2, 8]. Owing to the large band-gap, saturation velocity of charge carries and breakdown electrical field, AlGaN/GaN HFETs can operate at extremely high power density of tens of W/mm of the channel width, which is unattainable with other technologies [9-10]. Unfortunately, at such power levels, Joule heating starts to degrade performance of GaN devices causing reliability problems. The mean-time to failure (MTTF) of GaN transistors decreases rapidly with increasing junction temperature of the devices [1].

A number of methods have been used to improve heat removal from GaN devices. Conventional sapphire substrates with low thermal conductivity of $K$~30 W/mK at room temperature (RT) have been replaced with more expensive SiC substrates with the high thermal conductivity of $K$~100 – 350 W/mK at RT. However, even in GaN transistors on SiC substrate, self-heating can lead to temperature rises, $\Delta T$, above 180$^\circ$C. The composite substrates [5] and FC bonding [4] were utilized to improve the heat removal by reducing the thermal resistance on the scale of the whole wafer. Despite these efforts, the problems of the hot spots that develop near the downscaled device channels – at the nanometer and micrometer-scale – still persist. In this communication we demonstrate that thermal management of AlGaN/GaN heterostructure field-effect transistors (HFET) can be substantially improved via introduction of the additional heat-escaping channels – top-surface heat spreaders – made of few-layer graphene (FLG). FLG reveals an order-of-magnitude higher thermal conductivity, $K$, than that of GaN, which ranges from 125 to 225 W/mK at RT [6-7].

Here, we propose and examine a conceptually alternative approach for thermal management of high-power density devices. Instead of trying to further reduce the thermal resistance, $R_T$, of the substrate, we introduce the lateral heat spreaders on top of the GaN device structure, which provide additional heat escape channels from the hot spots. Ideally, the heat spreader should be made from material with the highest *K* as possible. It was recently discovered by some of us that graphene has the highest intrinsic thermal conductivity of all known materials, which increases with the lateral size [11-12]. From the practical applications point of view, FLG is better because its *K* is less subject to deterioration due to extrinsic effects, e.g. defects and disorder at the interfaces [12]. In addition, FLG and thin graphite films allow for the larger in-plane heat flux through its cross-section while still preserving graphene's mechanical flexibility.

**Experimental Results**

In order to perform the proof-of-concept demonstration, we transferred FLG and graphite films exfoliated from the highly-oriented pyrolytic graphite (HOPG) to the AlGaN/GaN devices on SiC substrate. Some of the flakes were naturally attached to thicker graphite regions. Graphite has high thermal conductivity ($K \approx 2000$ W/mK) and can be utilized both for heat spreaders and heat sinks depending on the thickness, geometry and size [12]. The fast progress in various chemical methods of large-area FLG growth [13-16], stimulated by strong interest to graphene, suggests that deposition of FLG and graphite on substrates with controlled number of the atomic planes, *n*, will soon become a commercial technology. The latter will facilitate practical applications of FLG for heat spreading. The graphite heat sinks can also be replaced with the metallic sinks or vertical thermal vias connected to the bottom heat sink.

Figure 1a-f illustrates the concept of "graphene-graphite quilts" as the top-surface heat spreaders and provides their microscopy images. We used AlGaN/GaN HFETs with the layered structure consisting of 30-nm AlGaN (~20% Al) barrier on 0.5-µm-thick GaN

channel layer deposited on insulating 4H-SiC substrate. The source and drain metal contacts were made of Ti/Al/Ti/Au while the gate electrode was made of Ni/Au. Details of the device structure and fabrication of typical AlGaN/GaN HFETs can be found in literature [8-10].

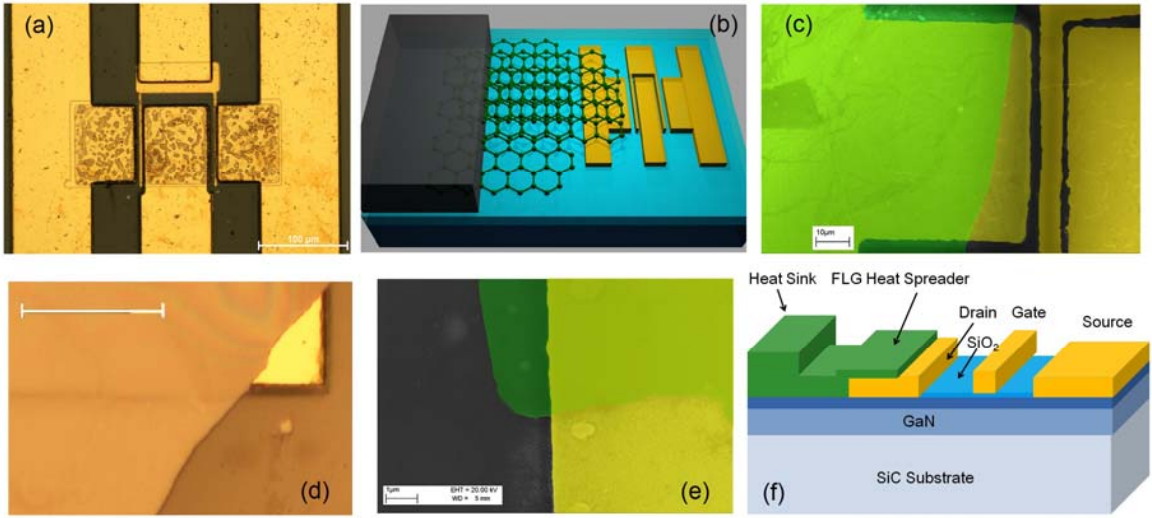

**Figure 1:** "Graphene-graphite quilts" as the top-surface heat spreaders for AlGaN/GaN HFETs. 1a: Optical microscopy of AlGaN/GaN HFETs before fabrication of the heat spreaders. 1b: Schematic of the FLG-graphite heat spreaders attached to the drain contact of the AlGaN/GaN HFET. 1c: SEM image of the heat spreader transferred to the drain contact. The "graphene-graphite quilt" is indicated with the green color while metal contacts are with the yellow color. 1d: Optical microscopy image of the "quilt" overlapping the metal drain contact and GaN surface demonstrating the heat spreader's flexibility and its close contact with the surface. 1e: SEM image of the heat spreader – metal contact region and GaN surface. 1f: Schematic of the device structure and the "graphene-graphite quilt" used in the simulation for the heat spreader optimization. Dark blue indicates the AlGaN barrier layer. Note that the FLG layer can be extended all the way to the space between the drain and the gate.

The gate length and widths of the devices used in this study were 3.5 μm and 90 μm, respectively. The large source-drain separation of 12 μm facilitated the heat spreader fabrication. The direct HOPG exfoliation on GaN/SiC substrate cannot be accomplished owing to the random nature of the process. For this reason, we applied the PMMA-assisted method [17] with some modifications, which allowed us an accurate placement of FLG-graphite "quilts" in pre-determined locations on top of GaN devices. The heat spreaders were attached to the drain contacts of the devices – the closest to the hot spots – according to reported simulations [18-19]. We carefully avoided short-circuiting GaN devices making sure that the "graphene quilts" extend from the drains directly to the graphite heat sinks on the side of the devices. Figure 2 shows the schematic of the process of FLG transfer to GaN structure. The details are given in the *Methods* section.

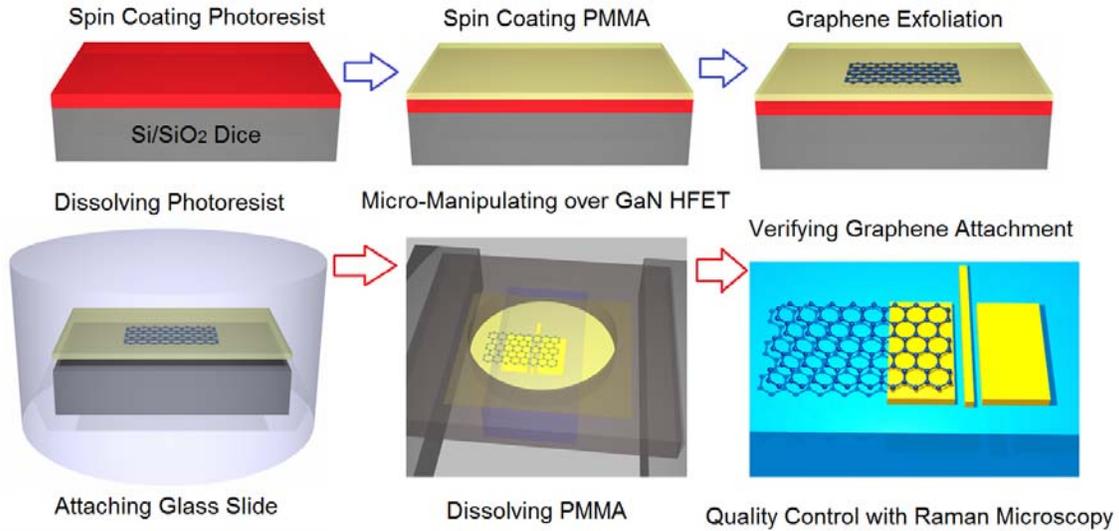

**Figure 2:** Transfer of FLG films to exact locations on AlGaA/GaN HFET device structure. The first set of steps, indicated by blue arrows, includes spin coating of the photoresist and PMMA, followed by FLG exfoliation on top of PMMA. The second set of steps, indicated by red arrows, consists of dissolving photoresist and attaching graphene to glass slides. The slide is micro-manipulated over AlGaN/GaN HFET to exact location under optical microscope. Dissolution of PMMA leaves FLG film in the desired location. Micro-Raman inspection confirms the number of layers and the quality of FLG after the transfer process.

The AlGaN/GaN HFETs with the heat spreaders and the reference HFETs without the heat spreaders have been wire-bonded and placed under the Raman microscope (Renishaw inVia). Raman spectroscopy was utilized both for quality control of FLG and graphite films after the transfer process and for *in-situ* monitoring of *ΔT* in the powered devices. The non-contact and non-destructive micro-Raman spectroscopy technique has been previously used for *T* mapping in AlGaN/GaN devices by measuring the temperature-dependent shifts in the Raman peak positions [10, 20]. We performed the measurements in the backscattering configuration under the 488-nm laser excitation. Figure 3a shows a typical spectrum from the FLG "quilt" on AlGaN/GaN/SiC device structure. The peaks at ~205 cm$^{-1}$, 610 cm$^{-1}$, 777 cm$^{-1}$ and 965 cm$^{-1}$ are the $E_2$ acoustic, $A_1$ longitudinal acoustic (LA), $E_2$ planar optical and $A_1$ longitudinal optical (LO) phonon modes of SiC [21]. The SiC $E_2$ peak at 777 cm$^{-1}$ was selected for monitoring $T$ rise in SiC substrate [22]. Owing to the finite penetration depth of the laser light the measured $\Delta T$ corresponds to the top part of SiC substrate – near the GaN channel layer. The temperature in the GaN channel itself was derived from the position of the narrow GaN $E_2$ (high) peak at 567 cm$^{-1}$.

The temperature coefficient of the SiC $E_2$ peak is -0.0144 cm$^{-1}$/K [22]. The $T$ dependence of the Raman peak position in GaN can be described as [23]

$$\omega(T) = \omega_0 - \alpha / [\exp(\beta hc\omega_0 / k_B T) - 1], \qquad (1)$$

where $\omega_0$ is the extrapolated Raman peak position at 0 K, $h$ is Plank's constant, $k_B$ is Boltzmann's constant, $c$ is the speed of light, $\alpha$ and $\beta$ are the fitting parameters [23-24]. The laser spot size was ~1 μm. Since the laser power on the sample surface was below ~2 mW and most of light went through GaN and SiC – the wide band gap semiconductors – the laser induced local heating was negligible compared to the power dissipated by the transistors. The experimental data were fitted with the Lorentzian functions to determine the Raman peak positions with the spectral resolution better than 0.1 cm$^{-1}$.

Figure 3b-c shows $E_2$ phonon peak in the Raman spectra of two AlGaN/GaN devices with and without the heat spreaders. The devices were located on the same wafer and had the same layered structure and dimensions. The laser spot was focused at the channel region between the gate and the drain, where $\Delta T$ is expected to be the highest. The source-drain bias, $V_{DS}$, was varied from 0 V to ~20 V with the 4-V intervals. The Raman peak position shifted to lower wavenumbers with increasing $V_{DS}$ indicating temperature rise with increasing dissipated power owing to the Joule heating in the channel. At the power density, $P$, of ~12.8

W/mm the temperature rise, $\Delta T$, was 92°C for the AlGaN/GaN HFET with the "graphene-graphite quilt" and $\Delta T$=118°C in the HFETs without the heat spreader. In this specific example, the same $P$ was achieved in the devices at 20 V and 22 V due to small variations in the current-voltage characteristics (I-Vs). The corresponding $\Delta T$ in the top region of SiC substrate was 44°C and 30°C in the HFETs without and with the heat spreader, respectively. Our experimental data indicate that even at the moderate $P$, the "graphene-graphite quilts" can help to reduce the hot-spot temperature in AlGaN/GaN HFETs.

Figures 3d and 3e present I-V characteristics of representative AlGaN/GaN HFETs while Figure 3f gives comparison of I-Vs of the HFETs with and without the heat spreaders. The tested device was completely pinched off at negative gate bias $V_G$=-4 V. The maximum source-drain current density of $I_{SD}$=0.75 A/mm was obtained at a positive gate bias $V_G$=2 V. The negative-slope regions in I-V curves indicate a degradation of the carrier mobility due to Joule heating as the dissipated power increases (see Figure 3d). Figure 3e shows I-Vs of the same device at elevated ambient temperature, $T_A$. The ambient temperature was changed by placing the device on a hot chuck. The saturation current density, $I_{DS}$, decreases rapidly with increasing $T_A$. For a given device $I_{DS}$ follows the equation $I_{DS} = 0.487 - 0.0014 \times T_A$. The obtained I-V characteristics and their dependence on $T_A$ are in agreement with literature [X-X]. The data illustrate the importance of temperature effects on performance of AlGaN/GaN HFETs. The insets to Figure 3e shows one of the tested devices and Raman SiC $E_2$ peak for the low and high $V_{DS}$ in HFET without the heat spreader.

Figure 3f provides direct comparison of I-Vs characteristics before (solid lines) and after (dashed lines) introduction of the "graphene-graphite quilts". At $V_G$=2 V, $I_{SD}$ increases from ~0.75 A/mm to ~0.84 A/mm – 12 % improvement – as a result of better heat removal with the top lateral heat spreaders. At $V_G$=0 V, $I_{SD}$ increased from 0.47 A/mm to 0.51 A/mm, which is 8 % improvement. At $V_G$=-2V, the current density remains almost the same after introduction of the heat spreader owing to the low dissipation power density at this negative gate bias. These experiments present a direct evidence of the improvement in the AlGaN/GaN HFET performance with the top-surface heat spreaders.

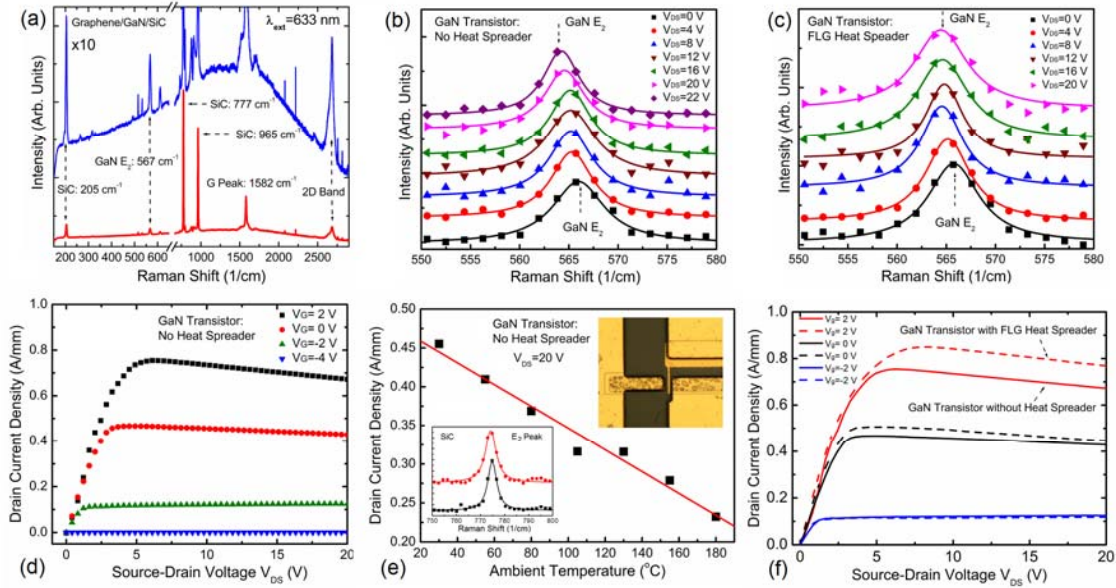

**Figure 3:** Micro-Raman monitoring of the temperature rise in the device structure and I-V characteristics of AlGaN/GaN HFETs. 3a: Raman spectrum of FLG on top of AlGaN/GaN structure on SiC substrate. The luminescence background alters the intensity ratios. To make the small peak easily distinguishable the original spectrum (red) is shown under ×10-magnification (blue). 3b: $E_2$ peak shift in AlGaN/GaN HFET without the heat spreader at the power density P=12.8 W/mm. 3c: Smaller $E_2$ peak shift in AlGaN/GaN HFET with the heat spreader indicating a reduction in the temperature rise at the same power density. 3d: I-V characteristics of a typical tested device showing self-heating effects at the high current density. 3e: Saturation current degradation with increasing ambient temperature. Insets show one of the tested devices and SiC $E_2$ Raman peak used for temperature monitoring inside the GaN substrate. 3f: Comparison of I-Vs of AlGaN/GaN HFETs with and without "graphene-graphite quilts" indicating improvement in I-Vs of HFETs with the heat spreaders.

**Simulation Results and Discussion**

To rationalize the experimental results and estimate the achievable improvements in thermal management of AlGaN/GaN HFETs on different substrates we simulated heat propagation in AlGaN/GaN layered device structures using the finite-element method (see *Supplementary Materials*). The thickness of SiC substrate, GaN, AlGaN and $SiO_2$ layers were 0.4 mm, 0.5 μm, 30 nm and 10 nm, respectively. Their corresponding RT thermal conductivities were taken from literature to be 350 W/mK, 160 W/mK, 120 W/mK and 1.4 W/mK [7-8, 25]. Thermal conductivity of FLG was assumed to be 2000 W/mK [12]. Thermal boundary resistance (TBR) at the GaN-substrate interface plays an important role in self-heating effects [18-19]. The TBR value $R_B=1.5\times10^{-8}$ m$^2$K/W was chosen to be consistent with the reported experimental [2, 26] and numerical studies [27]. To make sure that the assumed $K$ values for the layers are reasonable, we measured the effective thermal conductivity of the whole AlGaN/GaN/SiC device structure. The measurements were performed using the "laser-flash" technique (see *Supplementary Materials*). The affective $K$ for the whole structure was 300±56 W/mK at RT. This value is in line with the data from literature for each individual layer taking into account the unavoidable contributions of TBR at several interfaces [18].

In the model, we defined the heat source at the AlGaN-GaN interface and selected the boundary conditions at the substrate bottom to be RT. To validate the developed model with the experimental data we simulated $T$ distribution in AlGaN/GaN HFETs without the heat spreader at $P$=12.8 W/mm. The device structure parameters corresponded to the actual tested HFET. The simulation gave $\Delta T$=119°C for GaN channel (see Figure 4a), which is in excellent agreement with the measured $\Delta T$=119°C. The simulated $\Delta T$ in the upper region of SiC is also in line with the experimental data. The procedure was repeated for the AlGaN/GaN HFETs with the heat spreader of the geometry similar to the experimental structure. The simulated $\Delta T$=102°C in GaN channel is in agreement with the measurement within the ~10% uncertainty (Figure 4b).

Using the validated model we determined $T$ profiles in AlGaN/GaN HFETs with different heat spreader designs (Figure 4c-d). We used FLG with the number of atomic planes $n$=10 and the heat sink located at the distance, $D$, of 10 μm and 1 μm. Addition of the "graphene-graphite quilt" reduces temperature of the hot spot. The device structures with the closely located heat sinks offer stronger $\Delta T$ reduction (Figure 4d). In the practical designs, the nearby heat sinks attached to the top-surface heat spreaders can be implemented with the vertical thermal vias. The benefits of the "quilts" become more pronounced when one considers AlGaN/GaN HFETs on sapphire. Sapphire is a common substrate for AlGaN/GaN HFETs, which is less expensive than SiC but has lower $K$. Figure 4e-f shows the $T$ profile in AlGaN/GaN HFETs on sapphire without (4e) and with (4f) FLG heat spreader operating at $P$=3.3 W/mm. The data show that by introducing the "graphene-graphite quilt" with the heat sink at $D$=10 µm one can achieve a drastic 68$^o$C reduction in the hot-spot temperature. The experimentally observed and computationally predicted reduction in $\Delta T$ can lead to more than an order of magnitude increase in MTTF of GaN HFETs [1]. Our results open up a novel industry-scale [28] application for graphene, FLG and related sp$^2$ carbon materials, and can provide an additional impetus for further development of the nitride-based technology [29-30].

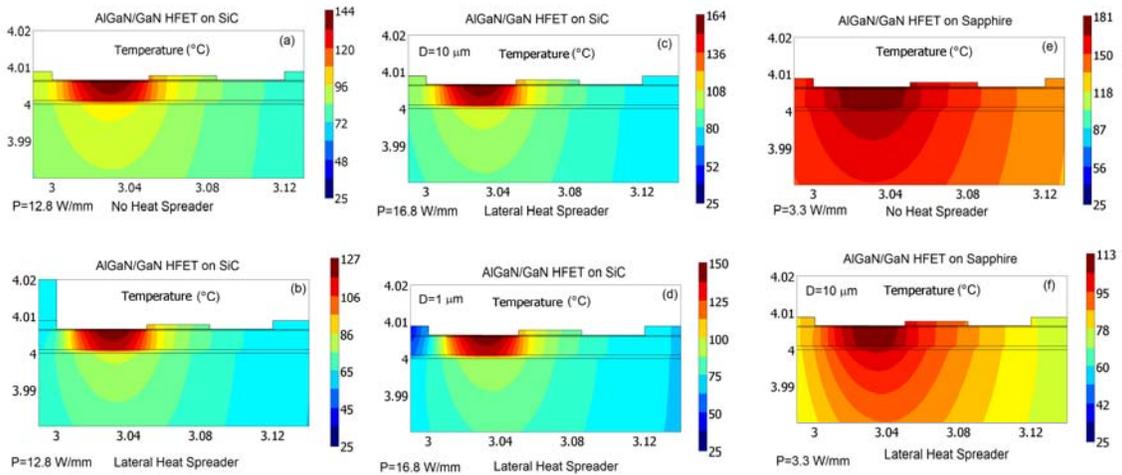

**Figure 4:** Simulated temperature distribution in AlGaN/GaN HFETs with different heat spreaders. 4a: Temperature distribution in AlGaN/GaN HFET without the heat spreader. 4b: Temperature distribution in the AlGaN/GaN HFET with the heat spreader, which has sizes matching one of the experimental structures. 4c-d: Temperature profile in the identical HFETs on SiC substrate powered at 16.8 W/mm with the heat spreader and the heat sink located at 10-µm distance (4c), and at 1-µm distance (4d). 4e-f: Temperature profile in HFETs on sapphire substrate powered at 3.3.W/mm without heat spreader (4e) and with the heat spreader (4f). The stronger effect is owing to the much lower thermal conductivity of sapphire. The HFET dimensions were kept the same in all simulations.

*METHODS*

**FLG transfer to GaN/SiC substrate:** It was required to transfer FLG films on top of AlGaN/GaN devices placing them precisely on the drain electrodes. The channel region near the drain generates most of heat according to our simulations. The FLG film should not touch other exposed electrodes and should be connected to larger graphite bars acting as the heat sinks. We adopted a method, which utilizes PMMA as the supporting membrane for transfer to the desired location [17], but modified it for our purposes. The innovation to the transfer procedure, which we introduced, is that we coat the substrate with PMMA before exfoliation of graphene on top of it. As a result, only one side of FLG contacts another material reducing possible residue. First, we spin coated a layer of photoresist (Shipley 1813) with 3500 rpm and baked it at 110°C for 90 sec. The substrate was then exposed with UV light. A second layer of polymer PMMA was spin coated with 3500 rpm and baked at 130°C for 90 sec. We followed with the mechanical exfoliation from HOPG to produce graphene on the substrate. After the coating procedure, FLG films (even the single layer graphene) can be optically identified under the microscope. Completing the procedure, we immersed the samples in photoresist developer (AZ400: H2O=1: 4) and dissolved the photoresist. The PMMA membranes were floated in the liquid. We used metal slides with the holes to attach the PMMA membranes. Note that the locations of the FLG films were roughly estimated to make sure that they fell into the holes. After removing the membranes from the liquid, the metal slides were mounted the on a micromanipulator, which was used for alignment. We were able to see the FLG films through the hole under the optical microscope and adjust the position of the substrate to place graphene on top of the desired location. The PMMA membranes were dissolved by hot acetone leaving FLG films attached to the substrate.

*Acknowledgements*
This work was supported by the US Office of Naval Research (ONR) through award No. 00014-10-1-0224 on Graphene Lateral Heat Spreaders for GaN Power Electronics. The authors thank Dr. Daniel S. Green for many illuminating discussions and Dr. Sergey Rumyantsev and Prof. Michal Shur for providing some of the AlGaN/GaN devices for this study. Special thanks go to Dr. Samia Subrina, Ms. Jie Yu and other NDL group members for the help with the simulations and measurements.

# References

[1] Trew, R. J., Green, D. S. & Shealy, J. B. AlGaN/GaN HFET reliability. *IEEE Microw. Mag.* **10,** 116 (2009).

[2] Kuzmik, J. *et al.* Self-heating phenomena in high-power III-N transistors and new thermal characterization methods developed within EU project TARGET. *Int. J. Microwave and Wireless Tech.* **1(2),** 153 (2009).

[3] Green, D. S. *et al.* GaN HEMT thermal behavior and implications for reliability testing and analysis. *Phys. Stat. Sol. (c)* **5,** 2026 (2008).

[4] Sun, J. *et al.* Thermal management of AlGaN–GaN HFETs on sapphire using flip-chip bonding with epoxy underfill. *IEEE Elec. Dev. Lett.* **24,** 375 (2003).

[5] Kidalov, S.V. & Shakhov, F. M. Thermal conductivity of diamond composites. *Materials* **2,** 2467 (2009).

[6] Zou, J., Kotchetkov, D., Balandin, A. A., Florescu, D. I. & Pollak, F. H. Thermal conductivity of GaN films: effects of impurities and dislocations. *J. Appl. Phys.* **92,** 2534 (2002).

[7] Liu, W. L. & Balandin, A. A. Thermal conduction in AlGaN alloys and thin films. *J. Appl. Phys.* **97,** 073710 (2005).

[8] Balandin, A. A. *et al.* Low flicker-noise GaN/A1GaN heterostructure field-effect transistors for microwave communications. *IEEE Trans. Microwave Theory and Techniq.,* **47,** 1413 (1999).

[9] Rajan, S. *et al.* Power performance of AlGaN–GaN HEMTs grown on SiC by plasma-assisted MBE. *IEEE Elec. Dev. Lett.,* **25,** 247 (2004).

[10] Kuball, M. *et al.,* Measurement of temperature in active high-power AlGaN/GaN HFETs using Raman spectroscopy. *IEEE Elec. Dev. Lett.* **23,** 7 (2002).

[11] Balandin, A. A. *et al.* Superior thermal conductivity of single-layer grapheme. *Nano Lett.,* **8,** 902 (2008).

[12] Balandin, A. A. Thermal properties of graphene and nanostructured carbon materials. *Nature Mat.,* **10,** 569 (2011).

[13] Eda, G., Fanchini, G. & Chhowalla, M. Large-area ultrathin films of reduced graphene oxide as a transparent and flexible electronic material. *Nature Nanotech.* **3,** 270–274 (2008).

[14] Park, S. & Ruoff, R. S. Chemical methods for the production of graphenes. *Nature Nanotech.* **4,** 217–224 (2009).

[15] Kim, K. S. *et al.* Large-scale pattern growth of graphene films for stretchable transparent electrodes. *Nature* **457,** 706–710 (2009).

[16] Obraztsov, A. N. Chemical vapour deposition making graphene on a large scale. *Nature Nanotech.* **4,** 212 (2009).

[17] Dean, C. R. *et al.* Boron nitride substrates for high-quality graphene electronics. *Nature Nanotech.* **5,** 722–726 (2010).

[18] Turin, V. O. & Balandin, A. A. Performance degradation of GaN field-effect transistors due to thermal boundary resistance at GaN/substrate interface. *Electron. Lett.* **40,** 81 (2004).

[19] Turin, V. O. & Balandin, A. A. Electro-thermal simulations of the self-heating effects in GaN-based field-effect transistors. *J. Appl. Phys.* **100,** 054501 (2006)

[20] Kuball, M. *et al.* Measurement of temperature distribution in multifinger AlGaN/GaN heterostructure field-effect transistors using micro-Raman spectroscopy. *Appl. Phys. Lett.* **82,** 124 (2003).

[21] Burton, J. C., Sun, L., Long, F. H., Feng, Z. C. & Ferguson, I. T. First- and second-order Raman scattering from semi-insulating 4*H*-SiC. *Phys, Rev. B* **59,** 7282 (1999).

[22] Bauer, M., Gigler, A. M., Huber, A. J., Hillenbrand, R. & Stark, R. W. Temperature-depending Raman line-shift of silicon carbide. *J. Raman Spec.,* **40,** 1867 (2009).

[23] Liu, M. S. *et al.* Temperature dependence of Raman scattering in single crystal GaN films. *Appl. Phys. Lett.* **74,** 3125 (1999).

[24] Sarua, A. *et al.* Integrated micro-Raman/infrared thermography probe for monitoring of self-heating in AlGaN/GaN transistor structures. *IEEE Trans. Electron Dev.* **53,** 2438 (2006)

[25] Sheppard. S. T. *et al.* High-power microwave GaN/AlGaN HEMT's on semi-insulating silicon carbide substrates. *IEEE Elec. Dev. Lett.* **20,** 161 (1999).

[26] Sarua, A. *et al.,* Thermal boundary resistance between GaN and substrate in AlGaN/GaN electronic devices. *IEEE Trans. Electron Dev.* **54,** 3152 (2007)

[27] Filippov, K. & Balandin, A. A. The effect of the thermal boundary resistance on self-heating of AlGaN/GaN HFETs. *MRS Internet J. Nitride Semiconductor Research* [Online], **8,** article 4 (2003); available at http://nsr.mij.mrs.org/8/4/

[28] Segal, M. Selling graphene by the ton. *Nature Nanotech.* **4,** 612 - 614 (2009).